\documentclass[aps,prb,reprint,amsmath,amssymb,floatfix,superscriptaddress]{revtex4-2}

\usepackage{graphicx,xcolor,multirow,bm,hyperref}

\usepackage[ignoreunlbld,norefs,nocites]{refcheck}

\begin{document}

\title{Correlation energy of the paramagnetic electron gas at the
  thermodynamic limit}

\author{Sam Azadi}

\email{sam.azadi@physics.ox.ac.uk}

\affiliation{Department of Physics, Clarendon Laboratory, University
  of Oxford, Parks Road, Oxford OX1 3PU, United Kingdom}

\author{N.\ D.\ Drummond}

\affiliation{Department of Physics, Lancaster University, Lancaster
  LA1 4YB, United Kingdom}

\author{Sam\ M.\ Vinko}

\affiliation{Department of Physics, Clarendon Laboratory, University
  of Oxford, Parks Road, Oxford OX1 3PU, United Kingdom}
\affiliation{Central Laser Facility, STFC Rutherford Appleton Laboratory,
Didcot OX11 0QX, United Kingdom}

\date{\today}

\begin{abstract}
The variational and diffusion quantum Monte Carlo methods are used to
calculate the correlation energy of the paramagnetic three-dimensional
homogeneous electron gas at intermediate to high density.  Ground
state energies in finite cells are determined using
Slater-Jastrow-backflow trial wave functions, and finite-size errors
are removed using twist-averaged boundary conditions and extrapolation
of the energy per particle to the thermodynamic limit of infinite
system size. Our correlation energies in the thermodynamic limit are
lower (i.e., more negative, and therefore more accurate according to
the variational principle) than previous results, and can be used for
the parameterization of density functionals to be applied to
high-density systems.
\end{abstract}

\maketitle

\section{Introduction}

The pairwise Coulomb repulsion between electrons results in many-body
correlations in electronic systems such as the homogeneous electron
gas (HEG)\@ \cite{Azadi22, Gino}. 
The so-called correlation energy is a negative correction
to the mean-field Hartree-Fock energy. Although the correlation energy
is usually only a small percentage of the total energy of an
electronic system, it is crucial for an accurate description of
chemical and electronic properties
\cite{Lowdin55,Jones15,Dreizler}. Unfortunately it is also the most
complicated part of the energy to calculate accurately.

The three-dimensional (3D) HEG plays a crucial role in our
understanding of the nature of electronic correlation in real
materials \cite{Pines66,Giuliani05, Loos16}. Moreover, the HEG is one
of the most important models for our understanding of bulk systems
under extreme conditions, such as warm dense matter, which is an
exotic, highly compressed state of matter that exists between solid
and plasma phases at high temperatures
\cite{Dornheim16,Groth17,Dornheim22}.  The correlation energy of the
3D HEG as a function of density \cite{Sun10,Loos11,Bhattarai18} is a
fundamental element in the description of the electronic properties of
real systems by density functional theory (DFT)
\cite{Hohenberg64,Kohn65}.  However, calculating the correlation
energy accurately requires many-body wave function-based methods
\cite{Gell57} such as quantum Monte Carlo (QMC) techniques
\cite{Ceperley80,Holzmann20, Ceperley77,Ceperley78,Azadi21,Spink13,Shepherd12,Ruggeri18}.
The variational (VMC) and diffusion quantum Monte
Carlo (DMC) methods \cite{Ceperley80,Foulkes01} are stochastic
approaches for obtaining expectation values of quantum operators.
These techniques are especially efficient for calculating the ground
state energies of interacting fermions. The main object is an
approximate trial wave function, whose accuracy governs the final
energy and intrinsic statistical fluctuations in the simulations.

The DMC simulations of Ceperley and Alder \cite{Ceperley80} presented
important data connecting the high- and low-density regimes of the
correlation energy of the 3D HEG\@. Their data have been used in
parameterizations of the correlation energy over a wide density range
and are frequently used in DFT calculations. Well-known
parameterizations that make use of Ceperley and Alder's results were
provided by Perdew and Zunger (PZ81) \cite{Perdew81}, Vosko, Wilk, and
Nusair (VWN80) \cite{Vosko80}, and Perdew and Wang (PW92)
\cite{Perdew92}, among others.  The PW92 functional includes five
parameters, two determined from analytic high-density constraints and
three by fitting to the QMC data.  A density parameter interpolation
(DPI) \cite{Sun10}, which was constructed by imposing four
high-density and three low-density constraints on a seven-parameter
functional form, provided a check based purely on the satisfaction of
the exact constraints. Spink \textit{et al.}\ \cite{Spink13} performed
QMC calculations for spin-unpolarized and spin-polarized 3D HEGs over
the high- and intermediate-density ranges, which can be regarded as
the most accurate QMC data reported so far. In the present work, we
provide new QMC data for the correlation energy of the paramagnetic
(i.e., spin unpolarized) 3D HEG, which are lower than previously
reported results.  We use long-range backflow correlations to make
fixed-node errors more consistent between different cell sizes. 
Instead of using the analytic finite-size corrections, 
we extrapolate our results to infinite system size which 
provides more accurate results at the thermodynamic limit
\cite{Azadi22}. QMC energies in finite simulation cells obey the
variational principle and it is reasonable to assume that the QMC
energy per particle extrapolated to infinite system size is also an
upper bound on the true energy per particle.  Hence the fact that our
energies are lower than previous works strongly suggests that our
results are more accurate.

We have used the VMC and DMC methods to obtain 3D HEG correlation
energies at different densities. In the VMC method, parameters in a
trial wave function are optimized according to the variational
principle, with energy expectation values calculated by Monte Carlo
integration in the $3N$-dimensional space of electron position
vectors.  In the DMC method, the imaginary-time Schr\"{o}dinger
equation is used to evolve a statistical ensemble of electronic
configurations towards the ground state. Fermionic antisymmetry is
maintained by the fixed-phase approximation, in which the complex
phase of the wave function is constrained to equal that of an
approximate trial wave function optimized within VMC\@.

The simplest fermionic wave function is a Slater determinant, which
describes exchange effects but not correlation. Multideterminant wave
functions and pairing (geminal) wave functions \cite{Marchi09} can
also be used. The most efficient method of going beyond the Slater
wave function is to multiply it by a Jastrow factor $\exp(J)$, resulting in a
Slater-Jastrow wave function \cite{Ceperley77,Ceperley78}.  The
Jastrow factor usually depends explicitly on the distances between
particles, introducing correlation into the wave function. The Jastrow
factor is positive everywhere and symmetric with respect to the
exchange of indistinguishable particles, so it does not change the
nodal surface defined by the rest of the wave function.  By evaluating
the orbitals in the Slater determinant at quasiparticle coordinates
${\bf X}$, which are functions of all the electron positions, we
introduce a backflow transformation \cite{Kwon98,Pablo06}, and the
resulting wave function is referred to as a Slater-Jastrow-backflow
(SJB) wave function.

\section{Trial wave function}

We used a SJB trial spatial wave function $\Psi({\bf R})=e^{J({\bf
    R})}S({\bf X}({\bf R}))$ for all the systems we have studied,
where ${\bf R}=({\bf r}_1,\ldots,{\bf r}_N)$ is the $3N$-dimensional
vector of electron coordinates. The antisymmetric Slater part $S$ is a
product of determinants of single-particle orbitals for spin-up and
spin-down electrons. The single-particle orbitals in $S$ are of the
free-electron form $\psi_\mathbf{k}(\mathbf{r})=\exp(i\mathbf{k} \cdot
\mathbf{r})$, where wavevector $\mathbf{k}$ is a reciprocal lattice
vector of the simulation cell offset by twist vector
$\mathbf{k}_\text{s}$, where $\mathbf{k}_\text{s}$ lies in the
supercell Brillouin zone. The Jastrow exponent, which is symmetric
under electron exchange, takes the form
\begin{eqnarray}
 J & = & U+P+H \nonumber \\ & = & \sum_{i<j} u(r_{ij}) + \sum_{i<j}
 p(\mathbf{r}_{ij}) +\sum_{i<j<k} h(r_{jk},r_{ik},r_{ij}), \nonumber
 \\
\end{eqnarray}
where
\begin{equation}
 u(r) = \sum_{l=0}^{N_u} \alpha_l r^l {(r - L_u)}^C \Theta(L_u -
 r), \label{eq:u}
\end{equation}
where $r$ is the minimum-image distance between two electrons, the
cutoff length $L_u$ is less than or equal to the radius of the largest
sphere that can be inscribed in the Wigner-Seitz cell of the
simulation cell, $C=3$ specifies how smooth the function is at the
cutoff length, $\Theta$ is the Heaviside step function, and
$\{\alpha_l\}$ are optimizable parameters, which differ for parallel-
and antiparallel-spin electrons. To satisfy the Kato cusp conditions
\cite{Kato, Pack}, we fix $\alpha_1=\Gamma/{(-L_u)}^C+\alpha_0 C/L_u$,
where $\Gamma=1/2$ for opposite-spin electrons and $\Gamma=1/4$ for
same-spin electrons. We chose $N_u=8$. The $p$ term has the symmetry
of the simulation-cell Bravais lattice and allows a description of
correlation in the ``corners'' of the simulation cell. Its form is
\begin{equation}
 p(\mathbf{r})=\sum_A a_A \sum_{\mathbf{G} \in A^+} \cos(\mathbf{G}
 \cdot \mathbf{r}), \label{eq:p_term}
\end{equation}
where $A$ represents a star of symmetry-equivalent, nonzero,
simulation-cell reciprocal-lattice vectors ${\bf G}$, and $A^+$ is a
subset of $A$ that consists of one out of each $\pm {\bf G}$ pair.
The $\{a_A\}$ are optimizable parameters.  We used 46 stars of ${\bf
  G}$ vectors in $p$.  The Jastrow also includes symmetric
three-electron terms \cite{Kwon93,Pablo12}
\begin{eqnarray}
h(r,r',r'') & = & \sum_{l=0}^{N_h} \sum_{m=0}^{N_h} \sum_{n=0}^{N_h}
c_{lmn}  r^l {(r')}^m {(r'')}^n \nonumber \\ && {} \times {(r-L_h)}^C
{(r'-L_h)}^C {(r''-L_h)}^C \nonumber \\&& {} \times \Theta(L_h-r)
\Theta(L_h-r') \Theta(L_h-r'') \nonumber \\
\end{eqnarray}
where $L_h$ is a cutoff length and $c_{lmn}$ are linear parameters.
Constraints were placed on the linear parameters to ensure that $h$ is
cuspless.  We chose $N_h=4$. Different $h$ terms, meaning different
$\{c_{lmn}\}$, may be used for electron triplets involving different
combinations of spins. However, in this work, the parameters in the
three-electron Jastrow factors were constrained to be independent of
spin.  

Including a backflow transformation in the trial wave function, the
Slater part of the wave function $S$ is evaluated at transformed
``quasiparticle'' coordinates ${\bf X}(\mathbf{R})=\mathbf{R}+{\bm
  \xi}(\mathbf{R})$, where
\begin{equation}
 {\bm \xi}_i(\mathbf{R})=\sum_{j\neq i} \eta(r_{ij}) \mathbf{r}_{ij} +
 \sum_{j\neq i} {\bm \pi}({\bf r}_{ij})
\end{equation}
is the backflow displacement of electron $i$. $\eta$ is a cuspless,
smoothly truncated, isotropic polynomial function of minimum-image
electron-electron distance $r_{ij}$. The polynomial coefficients are
optimizable parameters, and are different for parallel- and
antiparallel-spin electrons \cite{Pablo06}. The form of $\eta(r)$ is
mathematically equivalent to that of the Jastrow $u(r)$ term
[Eq.\ (\ref{eq:u}), with $\Gamma=0$ for same-spin electrons and
  optimizable for opposite-spin electrons].  Typically we used
$N_\eta=8$ in the polynomial expansions. The ${\bm \pi}$ term has the
form of the gradient of a Jastrow $p$ term [Eq.\ (\ref{eq:p_term})]:
\begin{equation}
{\bm \pi}({\bf r}) = -\sum_A c_A \sum_{{\bf G} \in A^+} \sin({\bf G}
\cdot {\bf r}) \, {\bf G},
\end{equation}
where the $c_A$ are optimizable parameters.  As the gradient of a
scalar field, the ${\bm \pi}$ term is irrotational. We used 44 stars
of ${\bf G}$ vectors in ${\bm \pi}$. The backflow parameters were
allowed to depend on the spins of the electron pairs.

The wave functions were optimized by variance minimization
\cite{Umrigar88,Neil05} followed by energy minimization
\cite{Toulouse}.  The \textsc{casino} package was used for all our QMC
calculations \cite{casino}.

\section{Finite-size effects}

Monte Carlo-sampled canonical ensemble twist-averaged (TA) boundary
conditions were used to reduce quasirandom single-particle finite-size
errors in total energies due to momentum quantization effects
\cite{Lin01,Azadi15,Drummond08,Holzmann16,Azadi19}. The Hartree-Fock
kinetic and exchange energies were used as control variates to improve
the precision of the twist-averaged energy.  Systematic finite-size
errors due to the use of the Ewald interaction rather than $1/r$ to
evaluate the interaction between each electron and its
exchange-correlation hole and the incomplete description of long-range
two-body correlations were removed by fitting $E(N)=E(\infty)+b/N$ to
the TA DMC energy per particle at different system sizes
\cite{Chiesa06}. Unlike the previous work of Spink \textit{et
  al.}\ \cite{Spink13}, we do not rely on analytic finite-size
correction formulas \cite{Chiesa06,Drummond08}, but instead use the
analytic results to provide the exponents used in finite-size
extrapolation formulas.  All our calculations were performed using
face-centered cubic simulation cells, maximizing the distance between
each particle and its closest periodic image.

At very high density $r_\text{s} \ll 1$, systematic finite-size
effects are more challenging.  In this regime, the QMC energy is close
to the Hartree-Fock energy, and hence the QMC energy per particle
initially shows the Hartree-Fock $O(N^{-2/3})$ scaling with system
size \cite{Drummond08}, before eventually crossing over to the
asymptotic $O(N^{-1})$ scaling when the finite-size error becomes
small compared with the correlation energy.

\section{Correlation energies}

We studied the paramagnetic 3D HEG at density parameters
$r_\text{s}=0.5$, 0.75, 1, 2, 3, 4, 5, 7, 10, and 20. For each
density, QMC calculations were performed for simulation cells with $N
= 130$, 226, and 338 electrons. Our DMC energies were extrapolated
linearly to zero time step $\tau$, with the target walker population
being varied in inverse proportion to the time step \cite{Suppl}.  The
difference between the twist-averaged DMC energy at small $r_\text{s}$
( i.e., $r_\text{s}\leq 1.0$)  obtained with time step $\tau =
0.02r_\text{s}^2$ and the energy at zero time step is not
statistically significant. The same behavior was observed at large
$r_\text{s}$ with $\tau = 0.01r_\text{s}^2$. The energies and
variances calculated using SJB wave functions for different system
sizes are reported in the Supplemental Material \cite{Suppl}. The
correlation energy is defined as the difference between the
Hartree-Fock energy per electron [which is
  $E_\text{HF}=3{(9\pi/4)}^{2/3}/(10r_\text{s}^2)-3{(9\pi/4)}^{1/3}/(4\pi
  r_\text{s})$ for the paramagnetic HEG] and the exact ground-state
energy per electron, where the latter is approximated by our SJB-DMC
results extrapolated to the limit of infinite system size.

\begin{figure}[htbp!]
 \centering
 \begin{tabular}{cc}
  \includegraphics[clip,scale=0.4]{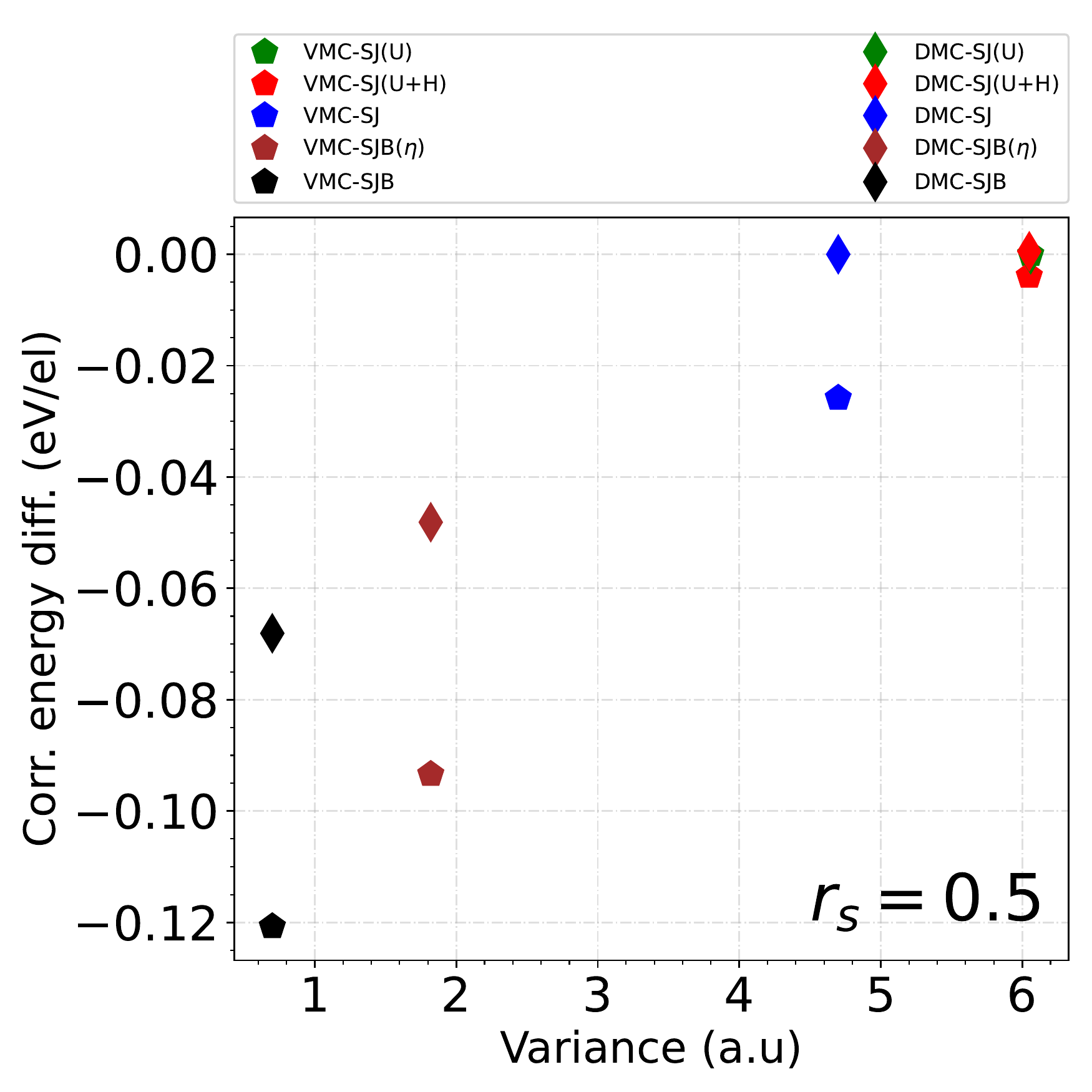}\\
  \includegraphics[clip,scale=0.4]{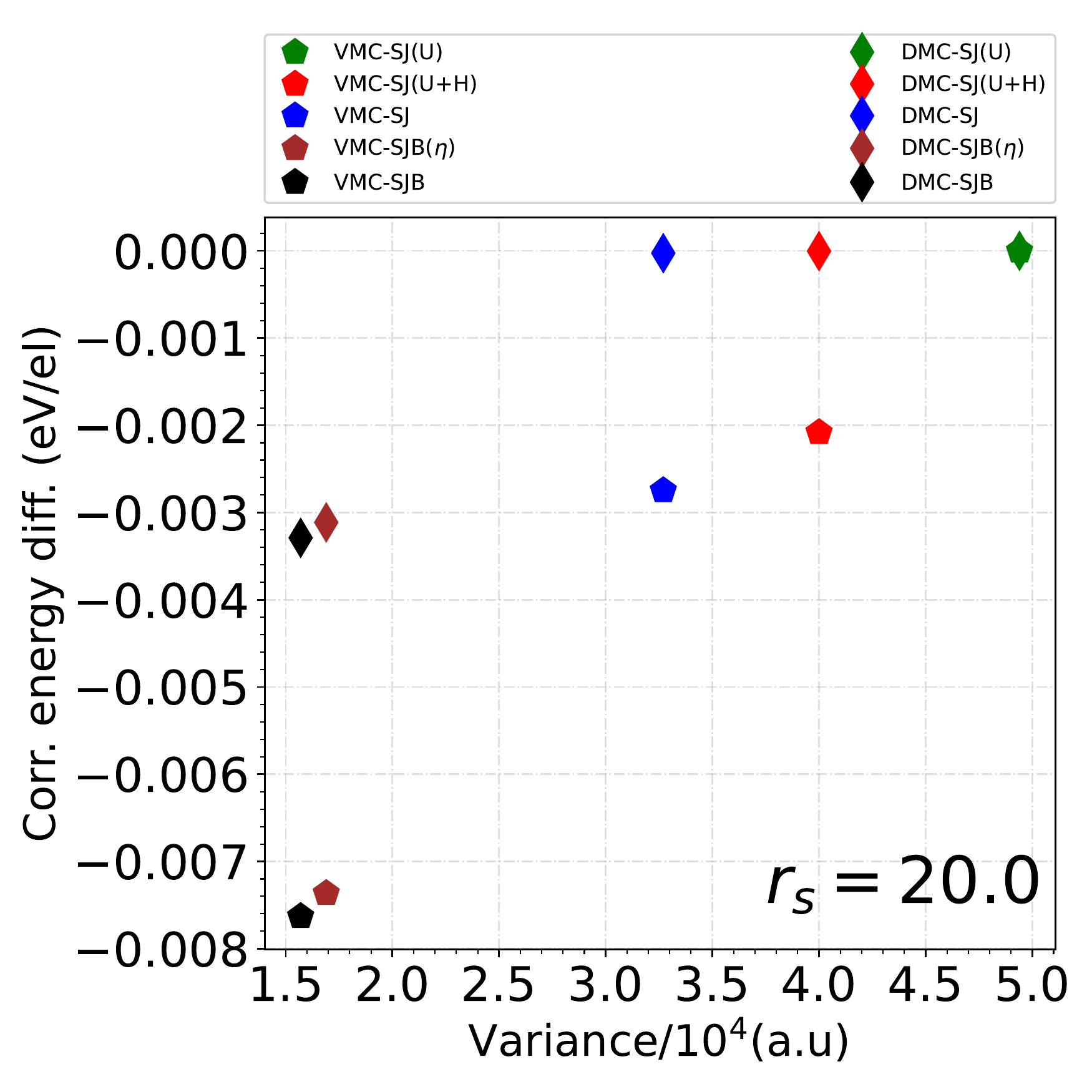}\\
 \end{tabular}
 \caption{\label{N54_Corr} Twist averaged VMC and DMC correlation
   energies at system size $N=54$ as a function of variance,
   relative to the correlation energy
   with a Slater-Jastrow (SJ) wave function in which the Jastrow
   factor only contains the isotropic two-body term $U$.  Results are
   shown at density parameters $r_\text{s}=0.5$ (top panel) and $20$
   (bottom panel). }
\end{figure}

Table~\ref{EN54} summarizes the contribution of each term of the trial
wave function to the correlation energy per particle in a simulation
cell containing $N=54$ electrons. We considered two systems with
$r_\text{s}=0.5$ and $r_\text{s}=20$. Figure~\ref{N54_Corr} shows the
improvements in the VMC and DMC correlation energies resulting from
the inclusion of different terms in the Jastrow and backflow
functions.

According to Spink \textit{et al.}\ \cite{Spink13}, the TA VMC energy
of the spin-unpolarized 3D HEG at $r_\text{s}=0.5$ at a system size of
$N=118$ is 3.41378(2) Ha/elec. Our TA VMC simulation for the same
system size yields the energy as 3.412460(4) Ha/elec., which is $\sim$
36 meV/elec.\ lower, because of the inclusion of the ${\bm \pi}$ term
in our work. 

\begingroup \squeezetable
\begin{table}[htbp!]
\centering
 \caption{\label{EN54} TA VMC and DMC correlation energies for $N=54$
   system size obtained using different terms in the Jastrow exponent
   and using both SJB and SJ wave functions.  Where Jastrow terms are
   not specified, $U$, $P$, and $H$ terms were used; where the
   backflow terms are not specified, $\eta$ and ${\bm \pi}$ terms were
   used.}
 \begin{tabular}{lcccc}
 \hline\hline

  \multirow{3}{*}{Wave fn} & \multicolumn{4}{c}{Correlation energy
    (eV/elec.)} \\

& \multicolumn{2}{c}{$r_\text{s}=0.5$} &
  \multicolumn{2}{c}{$r_\text{s}=20$} \\

& VMC & DMC & VMC & DMC \\

 \hline

  SJ($U$)   & $-2.5112(3)$ & $-2.5684(4)$ & $-0.311717(2)$ &
  $-0.317465(4)$ \\

  SJ($U+H$)   & $-2.5161(3)$ & $-2.5680(5)$ & $-0.313798(3)$ &
  $-0.317469(5)$ \\

  SJ   & $-2.5381(3)$ & $-2.5685(4)$ & $-0.314464(3)$ & 
  $-0.317492(5)$ \\

  SJB($\eta$) & $-2.6056(5)$ & $-2.6166(5)$ & $-0.319082(3)$ & 
  $-0.320579(5)$ \\

  SJB & $-2.6331(3)$ & $-2.6366(4)$ & $-0.319346(2)$ &
  $-0.320756(4)$ \\

 \hline\hline
 \end{tabular}
\end{table}
\endgroup

Our VMC and DMC energies extrapolated to the limit of infinite system
size are listed in Table~\ref{Evsrs}. To obtain the best linear fit
and reduce the noise in the energies as a function of the number of
particles $N$, a larger number of twists was used for
higher densities and smaller system sizes. The smallest and largest
number of twists were 120 and $10^4$, respectively. Comparing our
infinite-system VMC and DMC results with the DMC energies of Spink
\textit{et al.}\ \cite{Spink13} demonstrates not only the improvement
of the trial wave function due to the inclusion of long-range ${\bm
  \pi}$ backflow terms, but also the importance of removing
finite-size effects by extrapolation rather than relying on analytic
correction formulas.
\begingroup \squeezetable
\begin{table}[htbp!]
\centering
\caption{\label{Evsrs} TA VMC and DMC energies of the 3D HEG
  extrapolated to the thermodynamic limit from different system sizes
  ($N=130$, 226, and 338), compared with the DMC results of Spink
 \textit{et al.}\ \cite{Spink13}, and Ceperley and Alder \cite{Ceperley80}.
 Our DMC energies have been extrapolated to zero time step. }
 \begin{tabular}{lcccc}
 \hline\hline

 \multirow{3}{*}{$r_\text{s}$} & \multicolumn{4}{c}{Total energy
   (Ha/elec.)} \\

 & VMC & \multicolumn{3}{c}{DMC} \\

  & Pres.\ wk. & Pres.\ wk. & Spink \textit{et al.}  & Ceperley \& Alder \\

 \hline

  0.5  & $3.4255(1)$   & $3.42541(8)$   & $3.43011(4)$   & \dots\\

  0.75 & $1.28625(5)$   & $1.28620(5)$   & \dots         & \dots \\

  1.0  & $0.58643(4)$   & $0.58640(2)$   & $0.58780(1)$  & $0.5870(5)$  \\

  2.0  & $0.00195(2)$   & $0.001917(9)$  & $0.002380(5)$ & $0.002050(2)$ \\

  3.0  & $-0.06728(1)$  & $-0.067309(9)$ & $-0.067075(4)$  & \dots \\

  4.0  & $-0.07767(1)$  & $-0.07771(1)$  & \dots      & \dots\\

  5.0  & $-0.07594(1)$  & $-0.07597(1)$  & $-0.075881(1)$  & $-0.07560(5)$\\

  7.0  & $-0.066304(5)$ & $-0.066348(2)$ & \dots  & \dots  \\

  10.0 & $-0.053503(2)$ & $-0.053527(5)$ & $-0.0535116(5)$ & $-0.0533750(2)$ \\

  20.0 & $-0.031720(6)$ & $-0.031755(3)$ & $-0.0317686(5)$ & $-0.0316450(1)$\\

 \hline\hline
 \end{tabular}
\end{table}
\endgroup

Table~\ref{ECorr} compares our VMC and DMC results for the correlation
energy with the PZ81 \cite{Perdew81}, VWN80 \cite{Vosko80}, PW92
\cite{Perdew92}, and DPI \cite{Sun10} parameterizations, as well as
the DMC data of Spink \textit{et al.}  \cite{Spink13}.  Our
correlation energies are the lowest. Even at the high density
$r_\text{s}=0.5$ our DMC correlation energy is lower than the DPI
parameterization by $-16$ meV/elec.

\begin{figure}[htbp!]
    \centering
    \begin{tabular}{c}
     \includegraphics[clip,scale=0.8]{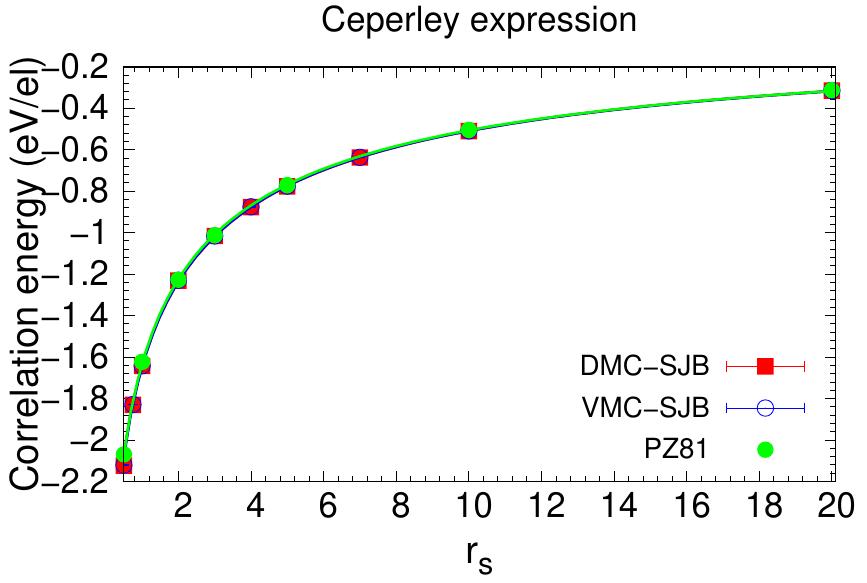}\\
    \includegraphics[clip,scale=0.8]{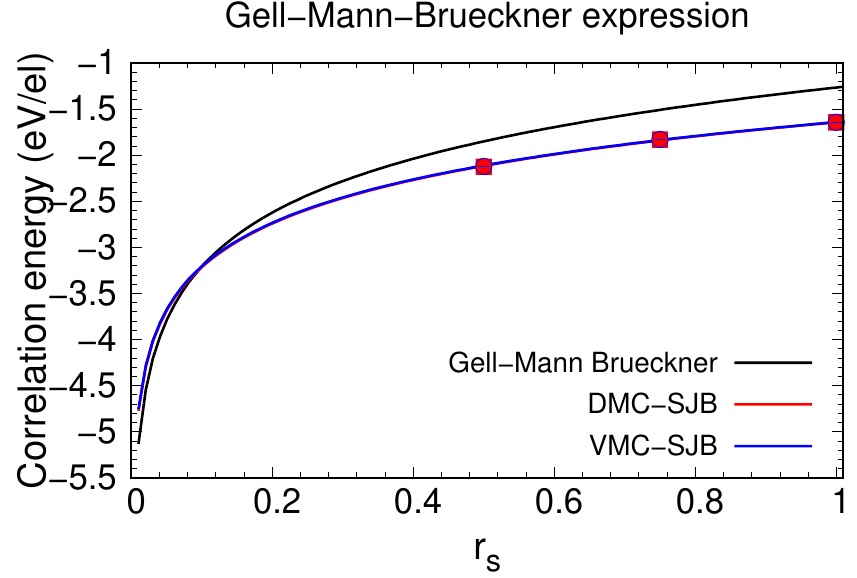}\\
    \includegraphics[clip,scale=0.8]{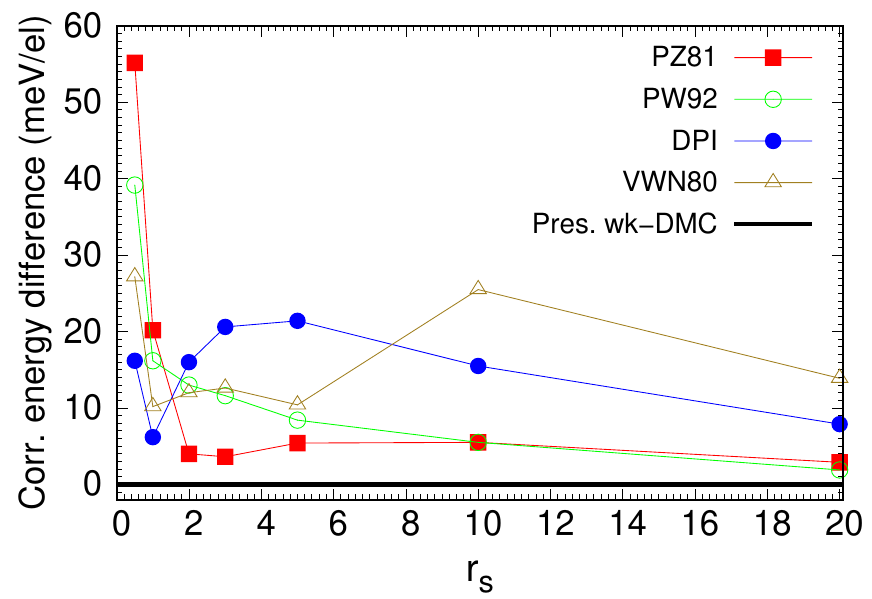}
    \end{tabular}
 \caption{\label{Corr_high} Correlation energy of the spin-unpolarized
   3D HEG as a function of $r_\text{s}$. (Top panel) The VMC and DMC
   correlation energies are fitted to Eq.\ (\ref{eq:ceperley}) and are
   compared with the PZ81 \cite{Perdew81} parameterization. (Middle
   panel) VMC and DMC correlation energies at high density
   ($r_\text{s}=0.5$, 0.75, and 1) are fitted to Gell-Mann and
   Brueckner's asymptotic formula. The fitting parameters are
   presented in the main text. (Bottom panel) Difference between our
   DMC correlation energy (Pres.\ wk-DMC) and the PZ81
   \cite{Perdew81}, VWN80 \cite{Vosko80}, PW92 \cite{Perdew92}, and
   DPI \cite{Sun10} DFT functionals.}
\end{figure}

Following Ceperley \cite{Ceperley78}, we fit
\begin{equation} 
 E_\text{c}(r_\text{s}) = \frac{\gamma}{1 + \beta_1\sqrt{r_\text{s}} +
   \beta_2 r_\text{s}} \label{eq:ceperley}
\end{equation}
to our SJB-DMC correlation energies (Fig.~\ref{Corr_high}).  The
fitting parameters $\gamma$, $\beta_1$, and $\beta_2$ are $-0.151(5)$
Ha $1.18(7)$, and $0.338(5)$, respectively.  The $\chi^2$ value of the
fit is 307.48 per degree of freedom. Equation
(\ref{eq:ceperley}) is accurate for large $r_\text{s}$, as we have
shown in our recent work on the low-density phase diagram of the HEG,
where our DMC correlation energies for $30 \leq r_\text{s} \leq 100$
were fitted to Eq.\ (\ref{eq:ceperley}) giving a $\chi^2$ per degree
of freedom of 0.521 \cite{Azadi22}. We found that the $\chi^2$
per degree of freedom becomes 0.698 by fitting DMC correlation energies for
$20 \leq r_\text{s} \leq 100$ to Eq.\ (\ref{eq:ceperley}) and we found
the fitting parameters $\gamma$, $\beta_1$, and $\beta_2$ to be
$-0.1278(55)$ Ha/elec., $0.897(53)$, and $0.299(12)$, respectively.
\begin{table*}[htbp!]
\centering
 \caption{\label{ECorr} Correlation energies for the spin-unpolarized
   3D HEG from the PZ81 \cite{Perdew81}, VWN80 \cite{Vosko80}, PW92
   \cite{Perdew92}, and DPI \cite{Sun10} parameters, DMC (Spink
   \textit{et al.}\ \cite{Spink13}), and this work (VMC and DMC)\@.}
 \begin{tabular}{lccccccc}
 \hline\hline

  \multirow{2}{*}{$r_\text{s}$} & \multicolumn{7}{c}{Correlation
    energy (eV/elec.)} \\

 & PZ81 & VWN80 & PW92 & DPI & DMC (Spink \textit{et al.}) & VMC & DMC
  \\

 \hline

  0.5  & $-2.069$ & $-2.097$ & $-2.085$ & $-2.108$ & $-1.996$ &
  $-2.121(3)$  & $-2.124(2)$ \\

  0.75 & \dots  & \dots  & \dots  & \dots  & \dots  & $-1.829(1)$  &
  $-1.831(1)$ \\

  1.0  & $-1.623$ & $-1.633$ & $-1.627$ & $-1.637$ & $-1.605$ &
  $-1.642(1)$  & $-1.6432(5)$ \\

  2.0  & $-1.227$ & $-1.219$ & $-1.218$ & $-1.215$ & $-1.218$ &
  $-1.2301(5)$ & $-1.2310(2)$ \\

  3.0  & $-1.013$ & $-1.004$ & $-1.005$ & $-0.996$ & $-1.010$ &
  $-1.0159(3)$ & $-1.0166(2)$ \\

  4.0  & \dots  & \dots  & \dots  & \dots  & \dots  & $-0.8758(3)$ &
  $-0.8769(3)$ \\

  5.0  & $-0.771$ & $-0.766$ & $-0.768$ & $-0.755$ & $-0.774$ &
  $-0.7756(3)$ & $-0.7764(3)$ \\

  7.0  & \dots  & \dots  & \dots  & \dots  & \dots  & $-0.6368(1)$ &
  $-0.6380(1)$ \\

  10.0 & $-0.505$ & $-0.485$ & $-0.505$ & $-0.495$ & $-0.510$ &
  $-0.5098(1)$ & $-0.5105(2)$ \\

  20.0 & $-0.313$ & $-0.302$ & $-0.314$ & $-0.308$ & $-0.316$ &
  $-0.3149(1)$ & $-0.3159(1)$ \\ 

 \hline\hline
 \end{tabular}
\end{table*}

According to the all-orders perturbation theory of Gell-Mann and
Brueckner \cite{Gell57} the correlation energy at high density is
given by $E_\text{c}(r_\text{s})= A\ln(r_\text{s}) + C +
O(r_\text{s}\ln(r_\text{s}))$, where $A = \frac{1}{\pi^2}[1-\ln(2)]
\approx 0.0311$ Ha and $C \approx -0.0465$ Ha. The appearance of powers
of $\ln(r_\text{s})$ in this formula shows that the correlation energy
is a nonanalytic function of $r_\text{s}$ for $r_\text{s} \rightarrow
0$ and describes the failure of the naive perturbation approach. The
constant term $C$ is the sum of the second order Onsager's exchange
integral and a numerical constant caused by the sum over divergent
contributions \cite{Onsager}. In practice, this asymptotic formula is
accurate only for very small $r_\text{s} \ll 1$.
We fitted our VMC and DMC correlation energies for $r_\text{s}=0.5$,
0.75, and 1 to the Gell-Mann-Brueckner expression
(Fig.~\ref{Corr_high}). The fitting parameters are
$A_\text{VMC}=0.0250(5)$ Ha, $C_\text{VMC}=-0.06030(2)$ Ha,
$A_\text{DMC}=0.0250(5)$ Ha, and $C_\text{DMC}=-0.06040(1)$ Ha.  The
difference between the VMC and DMC fitting parameters is small because
the random errors due to twist averaging at high density dominate.
Figure~\ref{Corr_high} shows that at $r_\text{s} \leq 0.1$ the
correlation energy predicted by the Gell-Mann-Brueckner formula
becomes smaller than VMC and DMC\@.

One can include an additional term in the Gell-Mann-Brueckner expansion and write
the high-density expansion of the correlation energy per electron as
\begin{equation}
E_\text{c}(r_\text{s}) = A\ln(r_\text{s}) + C + Br_\text{s}\ln(r_\text{s})+O(r_\text{s}), 
\label{eq:GellMan}
\end{equation}
where the exact value of the coefficient of
$r_\text{s}\ln(r_\text{s})$ is $B = 0.00922921$ \cite{Loos11}. We
fitted our DMC correlation energies for $r_\text{s} = 0.5$, 0.75, 1.0,
and 2.0 to the extended expansion, and we found that the fitting
parameters $A$, $C$, and $B$ are $0.0275(4)$, $-0.06001(7)$, and
$-0.0031(2)$ Ha/elec., respectively, with a $\chi^2$ value of 3.52579.

We fitted all our DMC correlation energies for $0.5 \leq r_\text{s}
\leq 100$, which are reported in this work and in our recently
published paper \cite{Azadi22}, to Eq.\ (\ref{eq:ceperley}) plus
Eq.\ (\ref{eq:GellMan}) and we found the $\chi^2$ per degree of
freedom to be 54.3. We searched for the best fit
with the smallest $\chi^2$ value and discovered that by adding a
$r_\text{s}^{-3/4}$ term to the sum of Eqs.\ (\ref{eq:ceperley}) and
(\ref{eq:GellMan}) the $\chi^2$ value is reduced to 1.26. Hence, the
DMC results indicate that the correlation energy
\begin{eqnarray}
E_\text{c}(r_\text{s}) & = & A\ln(r_\text{s}) + C + Br_\text{s}\ln(r_\text{s}) + \frac{D}{r_\text{s}^{3/4}} \nonumber \\ & & {} + \frac{\gamma}{1 + \beta_1\sqrt{r_\text{s}} + \beta_2 r_\text{s}},
\label{eq:correlation}
\end{eqnarray}
can describe the correlation energy of the 3D paramagnetic HEG within
the density range $0.5 \leq r_\text{s} \leq 100$, which covers the
high-, middle-, and low-density regimes. The fitting parameters are
listed in Table~\ref{Tab:FitParm}.  
\begin{table}[htbp!]
\centering
\caption{\label{Tab:FitParm} Fitting parameters of Eq.\ (\ref{eq:correlation}) in Hartree.}
 \begin{tabular}{lcc}
  \hline \hline
  Fitting parameter & Value &Asymptotic std.\ err. \\
   \hline 
  $A$ (Ha/elec.) & 0.000435098   & 0.0001665 \\
  $C$ (Ha/elec.) & $-0.00221852$   & 0.0008169\\
  $B$ (Ha/elec.) & $-3.02312 \times 10^{-7}$  & $1.493 \times 10^{-7}$\\
  $D$ (Ha/elec.) & $-0.0134875$    & 0.0006189\\
  $\gamma$ (Ha/elec.) & $-0.077337$  & 0.004517 \\
  $\beta_1$ & 0.470881   & 0.05071 \\
  $\beta_2$ & 0.262613   & 0.004956\\
   \hline \hline
 \end{tabular}
\end{table}

\section{Conclusion}

In conclusion, we have performed VMC and DMC simulations using SJB
trial wave functions to calculate the correlation energy of the
paramagnetic 3D HEG at high and intermediate densities. We corrected
finite-size errors by twist averaging and extrapolation to the
thermodynamic limit. Our correlation energies, which are lower than
previously reported results, can be used to parameterize the
correlation energy of the 3D HEG for use in DFT calculations.

\begin{acknowledgments}
S.A.\ and S.M.V.\ acknowledge support from the UK EPSRC grants EP/P015794/1 and EP/W010097/1, the Royal Society, and PRACE (Partnership for Advanced Computing in Europe) for awarding us access to the High-Performance Computing Center Stuttgart, Germany, through the Project No.\ 2020235573.
\end{acknowledgments}

\bibliography{PRBCorrEnerRevised_arxiv}

\end{document}